\documentclass[prb,showpacs,twocolumn]{revtex4}

\usepackage{epsfig}

\usepackage{graphics}

\usepackage{graphicx}

\usepackage{dcolumn}

\usepackage{bm}

\newcommand{\sba}{\begin{subeqnarray}}

\newcommand{\sea}{\end{subeqnarray}}

\def\cm-1{cm$^{-1}$}

\begin{document}

\title{Diagrammatic theory for Anderson Impurity Model.\\Stationary property of the thermodynamic potential}
\author{V.\ A.\ Moskalenko$^{1,2}$}
\email{v_moscalenco@yahoo.com}
\author{P.\ Entel$^{3}$}
\author{L.\ A.\ Dohotaru$^{4}$}
\author{R.\ Citro$^{5}$}
\affiliation{$^{1}$Institute of Applied Physics, Moldova Academy
of Sciences, Chisinau 2028, Moldova}
\affiliation{$^{2}$BLTP, Joint Institute for Nuclear Research, 141980 Dubna, Russia}
\affiliation{$^{3}$University of Duisburg-Essen, 47048 Duisburg, Germany} \affiliation{$^{4}$Technical University, Chisinau 2004, Moldova} \affiliation{$^{5}$Dipartimento di Fisica E.\ R.\ Caianiello,
Universit\'{a} degli Studi di Salerno and CNISM, Unit\'{a} di ricerca di Salerno, Via S. Allende, 84081 Baronissi (SA), Italy}
\date{\today}

\begin{abstract}


%

A diagrammatic theory around atomic limit is proposed for normal
state of Anderson Impurity Model. The new diagram method is based
on the ordinary Wick's theorem for conduction electrons and a
generalized Wick's theorem for strongly correlated impurity
electrons. This last theorem coincides with the definition of Kubo
cumulants. For the mean value of the evolution operator a linked
cluster theorem is proved and a Dyson's type equations for
one-particle propagators are established. The main element of
these equations is the correlation function which contains the
spin, charge and pairing fluctuations of the system. The
thermodynamic potential of the system is expressed through
one-particle renormalized Green's functions and the correlation
function. The stationary property of the thermodynamic potential
is established with respect to the changes of correlation
function.
\end{abstract}

\pacs{71.27.+a, 71.10.Fd} \maketitle

\section{Introduction}

The study of strongly correlated electron systems has become in
the last decade one of the most active fields of condensed matter
physics. One of the most important models of strongly correlated
electrons is the Anderson impurity model (AIM)$^{[1]}$. It is a
model of the system of free conduction electrons that interact
with the system of the electrons of the $d-$ or $f-$ shells of
impurity atoms. The impurity electrons are strongly correlated
because of strong Coulomb repulsion and they undergo the
hybridization with conduction electrons. This model has been
largely applied to heavy fermion systems where the local impurity
orbital is $f-$ orbital $^{[2]}$.

The interest towards the Anderson impurity model has also
increased with the advent of Dynamical Mean Field Theory (DMFT)
within which an infinite dimensional lattice models can be mapped
onto effective impurity models by using self-consistency
conditions$^{[3,4]}$. This model has also been intensively
investigated by using the method of equation of motion for
retarded and advanced Green's functions proposed by Bogoliubov and
Tiablikov $^{[5]}$ and developed in papers $^{[6,7]}$.

We propose a diagrammatic method to treat the AIM which is an
alternative approach to the equations of motion and
renormalization methods$^{[8,9]}$. The first attempt to develop
the diagrammatic theory for this problem was realized in the
paper$^{[10]}$. These authors used the expansion by cumulants for
averages of products of Hubbard transfer operators and their
algebra. Afterwards, other diagrammatic techniques dealing with
perturbation in the hybridization amplitude while keeping the
on-site correlation exact have been developed$^{[11-13]}$. Here we
use the thermodynamic perturbation theory and Matsubara
$^{[14,15]}$ Green's functions, considering the hybridization of
both groups of electrons as a perturbation.

The Hamiltonian of the model is written as
%
\begin{eqnarray}
H&=&H_{0}+H_{int},  \nonumber \\
H_{0}&=&H_{0}^{c}+H_{0}^{f},  \nonumber \\
H_{0}^{c}&=&\sum\limits_{\mathbf{k}\sigma }\epsilon (\mathbf{k})\
C_{\mathbf{k}\sigma }^{+}C_{\mathbf{k}\sigma },  \nonumber \\
H_{0}^{f}&=&\epsilon _{f}\sum\limits_{\sigma }f_{\sigma }^{+}f_{\sigma
}+Un_{\uparrow }^{f}n_{\downarrow }^{f}, \\
H_{int}&=&\frac{1}{\sqrt{N}}\sum\limits_{\mathbf{k}\sigma }\left(
V_{ \mathbf{k}\sigma }f_{\sigma }^{+}C_{\mathbf{k}\sigma
}+V_{\mathbf{k}\sigma
}^{\ast }C_{\mathbf{k}\sigma }^{+}f_{\sigma }\right) ,  \nonumber \\
n_{\sigma }^{f}&=&f_{\sigma }^{+}f_{\sigma },  \nonumber
\label{1}
%
\end{eqnarray}
where $C_{\mathbf{k}\sigma }(C_{\mathbf{k}\sigma }^{+})$ and
$f_{\sigma }(f_{\sigma }^{+})$ - annihilation (creation) operators
of conduction and impurity electrons with spin $\sigma $\
correspondingly, $\epsilon (\mathbf{k})$ is the kinetic energy of
the conduction band state $(\mathbf{k},\sigma )$, $\epsilon _{f}\
$ is the local energy of $f$  electrons, $U$ is the on-site
Coulomb repulsion of the impurity electrons and $N$ the number of
lattice sites. Both energies are evaluated with respect to the
chemical potential $\mu$ of the system. The perturbation $H_{int}$
is the hybridization interaction between conduction and localized
electrons. The Coulomb repulsion between impurity electrons is far
to large to be treated as perturbation and it must be included in
the main part of Hamiltonian $H_{0}$. The existence of this term
invalidates the Wick theorem for local electrons. Therefore first
of all we formulate the Generalized Wick theorem (GWT) for local
electrons preserving the ordinary Wick theorem for conduction
electrons. Our (GWT) really is the identity which determines the
irreducible Green's functions or the Kubo cumulants.

For example the chronological product of four local operators
averaged with respect to zero-order density matrix of electrons
has the form $^{[16]}$:
%
\begin{widetext}
%
\begin{eqnarray}
%
\left\langle
Tf_{1}f_{2}\overline{f}_{3}\overline{f}_{4}\right\rangle _{0}& = &
\left\langle Tf_{1}\overline{f}_{4}\right\rangle _{0}\left\langle
Tf_{2}
\overline{f}_{3}\right\rangle _{0} -\left\langle Tf_{1}\overline{f}
_{3}\right\rangle _{0}\left\langle
Tf_{2}\overline{f}_{4}\right\rangle _{0}+g_{2}^{(0)ir}[1,2|3,4].\label{2}
%
\end{eqnarray}
%
\end{widetext}
Here the symbol $\left\langle ...\right\rangle _{0}$ means
thermodynamical average on zero order distribution function. The
right -hand part of equation (2) contains the first two terms
which are the ordinary Wick contributions and last one named by us
as irreducible Green's function or Kubo cumulant which contains
spin, charge and pairing fluctuations of localized electrons. In
case when the statistical average of operators contains 6
operators $\left\langle Tf_{1}f_{2}f_{3}
\overline{f}_{4}\overline{f}_{5} \overline{f}_{6}\right\rangle
_{0}$, the right-hand part contains $3!$ terms with product of
three one-particle Green's functions, then there are nine terms in
the form of product of one-particle Green's function and
two-particle irreducible function. At last there is a
three-particle irreducible Green's function
$g_{3}^{(0)ir}[1,2,3|4,5,6]$. The Green's functions
$g_{n}^{(0)ir}[1,...,n\mid n+1,...,2n]$ which appear at the $n$-th
order of the perturbation determine the structure of the new
diagrammatic technique. Such definitions of irreducible Green's
functions has been used already for discussing the properties of
the Hubbard and other strongly correlated models $^{[17-21]}$.

The zero order Hamiltonian of the localized electrons can be
diagonalized by using the Hubbard transfer operators $\chi ^{mn}$
$^{[22]}$. By using these operators it is possible to calculate
the simplest irreducible Green's functions.


\section{Diagrammatical theory}


The full one-particle Matsubara Green's function of localized
electrons in interaction representation has the form:
\begin{eqnarray}
%
g_{\sigma \sigma^{\prime}}(\tau-\tau ^{\prime })&=&-\
\left\langle Tf_{\sigma }(\tau )\overline{f}_{\sigma ^{\prime
}}(\tau ^{\prime })U(\beta )\right\rangle _{0}^{c},\label{3}
%
\end{eqnarray}
%
where the index $c$ means connected diagrams. The operators are taken in the
interaction representation , $T$ is chronological operator.

For conduction electrons it is convenient to define the local operator
%
\begin{eqnarray}
%
b_{\sigma }&=&\frac{1}{\sqrt{N}}\sum\limits_{\mathbf{k}}
V_{ \mathbf{k}}C_{\mathbf{k}\sigma}.\label{4}
%
\end{eqnarray}
%
The corresponding full conduction electron Green's function  has the form
%
\begin{eqnarray}
%
G_{\sigma \sigma^{\prime}}(\tau-\tau ^{\prime })&=&-\
\left\langle Tb_{\sigma }(\tau )\overline{b}_{\sigma ^{\prime
}}(\tau ^{\prime })U(\beta )\right\rangle _{0}^{c},\label{5}
%
\end{eqnarray}
%
where $U(\beta)$ is the evolution operator
%
\begin{eqnarray}
%
U(\beta ) & = & T\exp (-\int\limits_{0}^{\beta }H_{int}(\tau
)d\tau ).\label{6}
%
\end{eqnarray}
%

Because the matrix element of hybridization $V_{ \mathbf{k}}$ is
absorbed by local operator $b_{\sigma }$ it is convenient to
introduce a new parameter $\lambda$,  which will be associated to
each vertex of the diagrams.  In such a way the order of
perturbation theory will be determined by $\lambda$ and not by the
matrix element $V_{ \mathbf{k}}$ of hybridization which can be
present even in zero order Green's function. In the last stage of
the calculation $\lambda$ will be put equal to one.

In zero order of
perturbation theory the Fourier representation of these functions
are ($\overline{\sigma }=-\sigma $):
%
\begin{eqnarray}
%
g_{\sigma \sigma^{\prime}}^0(i\omega )&=&\delta_{\sigma \sigma^{\prime}}
g_{\sigma }^{0}(i\omega ),\nonumber \\
g_{\sigma }^{0}(i\omega )&=&\frac{1-n_{\overline{\sigma }}}{i\omega
-\epsilon _{f}}+\frac{n_{\overline{\sigma }}}{i\omega -\epsilon _{f}-U},
\nonumber \\
n_{\overline{\sigma }}&=&\frac{\exp (-\beta \epsilon _{f})+\exp \left[
-\beta (2\epsilon _{f}+U\right] }{Z_{0}}, \\
Z_{0}&=&1+2\exp (-\beta \epsilon _{f})+\exp \left[ -\beta
(2\epsilon _{f}+U
\right] .\nonumber\label{7}\
%
\end{eqnarray}
%
Here $\omega \equiv\omega_{n} =(2n+1)\pi /\beta $ are the Matsubara odd
frequencies.

For the conduction electrons we have
%
\begin{eqnarray}
%
G_{\sigma \sigma^{\prime}}^0(i\omega )&=&\delta_{\sigma \sigma^{\prime}}
G_{\sigma }^{0}(i\omega ),\nonumber \\
G_{\sigma }^{0}(i\omega )&=&\frac{1}{N}\sum\limits_{\mathbf{k}}
\frac{|V_{\mathbf{k}}|^{2}}{i\omega -\epsilon
(\mathbf{k})}.  \label{8}
%
\end{eqnarray}
%

The presence in the definition of zero order Green's function
$G_{\sigma }^{0}$ of the square of matrix element of hybridization is the
consequence of our equation (4) and not of the perturbation.

The thermodynamical perturbation theory gives us the results for
one-particle Green's functions presented on the Fig.1 and Fig.2.
The double solid and dashed lines depict the renormalized and the
thin lines the bar propagators of conduction and impurity
electrons. The lines connect the crosses which depict the impurity
states. To crosses are attached two arrows one of which is ingoing
and one outgoing. They depict the annihilation and creation of the
electrons correspondingly. The crosses are the vertices of the
diagrams and a $\lambda$ multiplier is attached to each of them.
The index $n$ means $(\sigma _{n},\tau _{n})$. The summation on
the index $\sigma _{n}$  and the integration on the $\tau _{n}$
are intended. The rectangles with $2n$ indices and crosses depict
the irreducible $g_{n}^{(0)ir}[1,...,n\mid n+1,...,2n]$ Green's
functions. The sign of diagrams is determined by the parity (even
or odd) of the permutation of the Fermi operators necessary to
obtain the diagram.
%
\begin{figure*}[t]
%
\centering
\includegraphics[width=0.85\textwidth,clip]{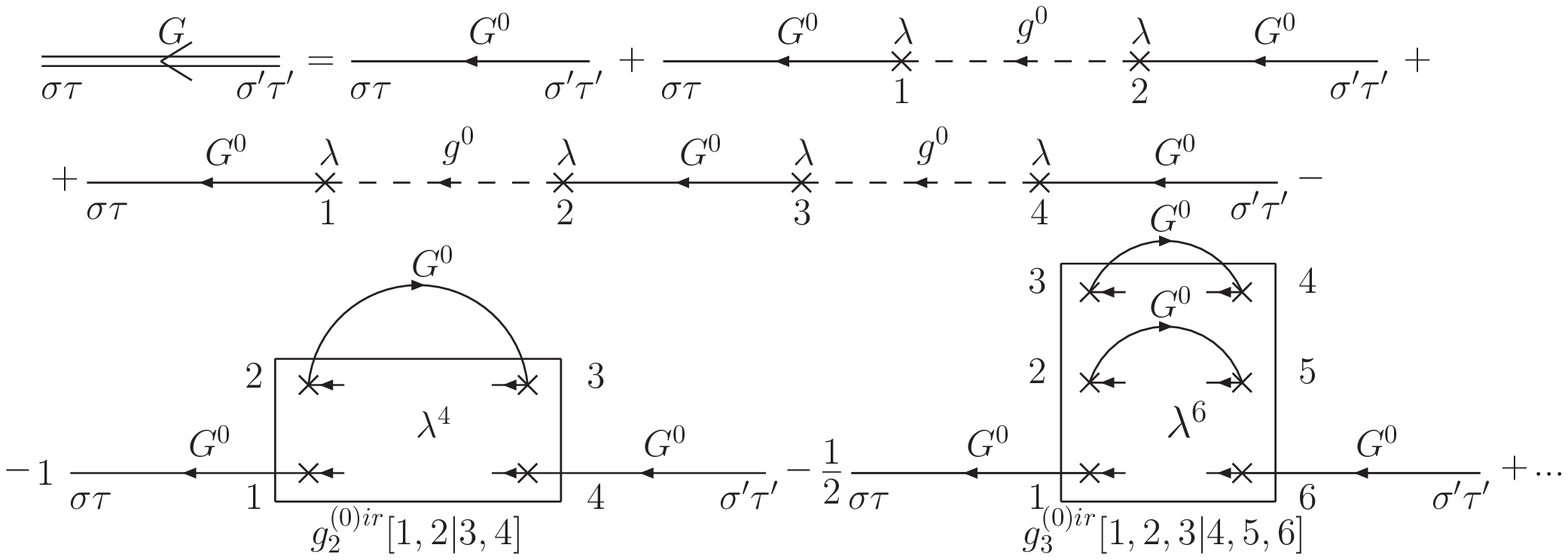}
\vspace{-0mm}
%
\caption{ Diagrams of first orders of perturbation theory for the
conduction electron propagator. The thin solid lines represent the
conduction electron propagator and dashed lines
the impurity electron propagator of zero order. Double line
represents the full
propagator. }\label{fig-1} \vspace{-5mm}
\end{figure*}
%
\begin{figure*}[t]
%
\centering
\includegraphics[width=0.85\textwidth,clip]{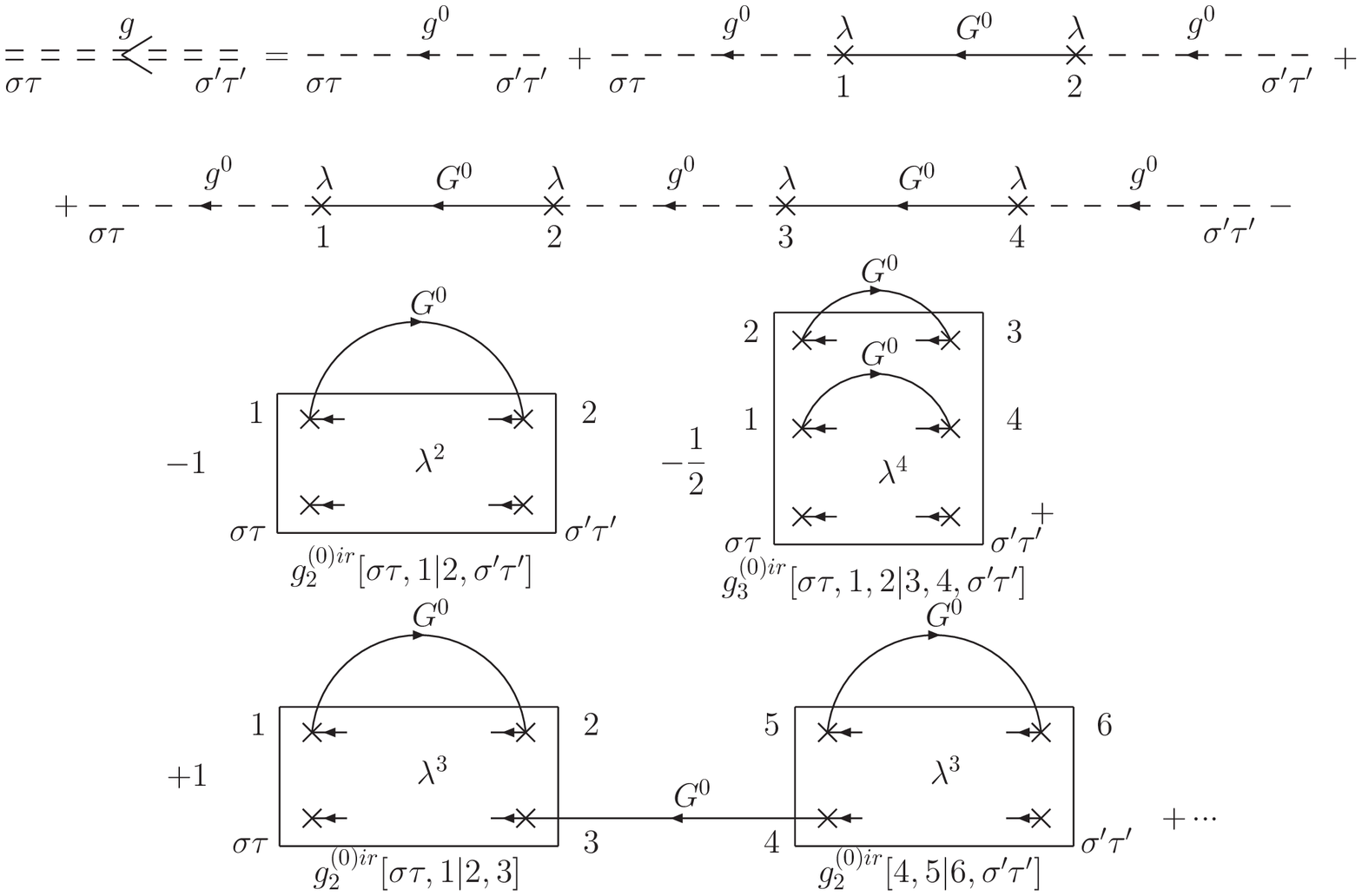}
\caption{ The diagrams for impurity electron propagator
$g_{\sigma \sigma^{\prime }}(\tau -\tau ^{\prime })$.
The last three diagrams contain the correlation contributions.
Two of them are strong connected and the last is weak
connected. }\label{fig-2} \vspace{-5mm}
%
\end{figure*}
Using Feynman's rules and the correspondence above, it is possible
to establish the next equation for the diagrams shown in Fig.1:
%
\begin{widetext}
%
\begin{eqnarray}
%
G_{\sigma \sigma ^{\prime }}(\tau -\tau ^{^{\prime }}|\lambda) & = &
G_{\sigma \sigma ^{\prime }}^0(\tau -\tau ^{^{\prime }})+\sum\limits_{\sigma _{1}\sigma_{2}}\int\limits_{0}^{\beta }d\tau _{1}\int\limits_{0}^{\beta
}d\tau _{2}G_{\sigma \sigma _{1}}^0(\tau -\tau _{1})\lambda g_{\sigma _{1}\sigma _{2}}(\tau _{1} -\tau _{2}|\lambda)\lambda G_{\sigma _{2} \sigma ^{\prime }}^0(\tau _{2} -\tau ^{^{\prime }}).\label{9}
%
\end{eqnarray}
%
\end{widetext}

On the basis of the diagrams depicted on the Fig.2, instead it is
possible to establish the following Dyson's type equation for
$g_{\sigma \sigma^{\prime }}$:
%
\begin{widetext}
%
\begin{eqnarray}
%
g_{\sigma \sigma ^{\prime }}(\tau -\tau ^{^{\prime }}|\lambda) & = &
\Lambda_{\sigma \sigma ^{\prime }}(\tau -\tau ^{^{\prime }}|\lambda)+\sum\limits_{\sigma _{1}\sigma_{2}}\int\limits_{0}^{\beta }d\tau _{1}\int\limits_{0}^{\beta
}d\tau _{2}\Lambda_{\sigma \sigma _{1}}(\tau -\tau _{1})\lambda
G_{\sigma _{1} \sigma _{2}}^0(\tau _{1} -\tau _{2})
\lambda g_{\sigma _{2} \sigma ^{\prime }}(\tau _{2} -\tau ^{^{\prime }}),\label{10}
%
\end{eqnarray}
%
\end{widetext}
where
%
\begin{eqnarray}
%
\Lambda_{\sigma \sigma ^{\prime }}(\tau -\tau ^{^{\prime }}|\lambda)& = &
g_{\sigma \sigma ^{\prime }}^0(\tau -\tau ^{^{\prime }})
+Z_{\sigma \sigma ^{\prime }}(\tau -\tau ^{^{\prime }}|\lambda).\label{11}
%
\end{eqnarray}
%
Here $Z_{\sigma \sigma ^{\prime }}$ is the new correlation
function which contains an infinite sum of the irreducible
Green's functions. As it was underlined above this
function contains all spin, charge and pairing fluctuations and is
the main element of our diagram technique.

Diagram representation of the correlation function
$\Lambda_{\sigma \sigma ^{\prime }}(\tau -\tau ^{^{\prime
}}|\lambda)$ is depicted on the  Fig.3
%

\begin{figure*}[t]
%
\centering
\includegraphics[width=0.85\textwidth,clip]{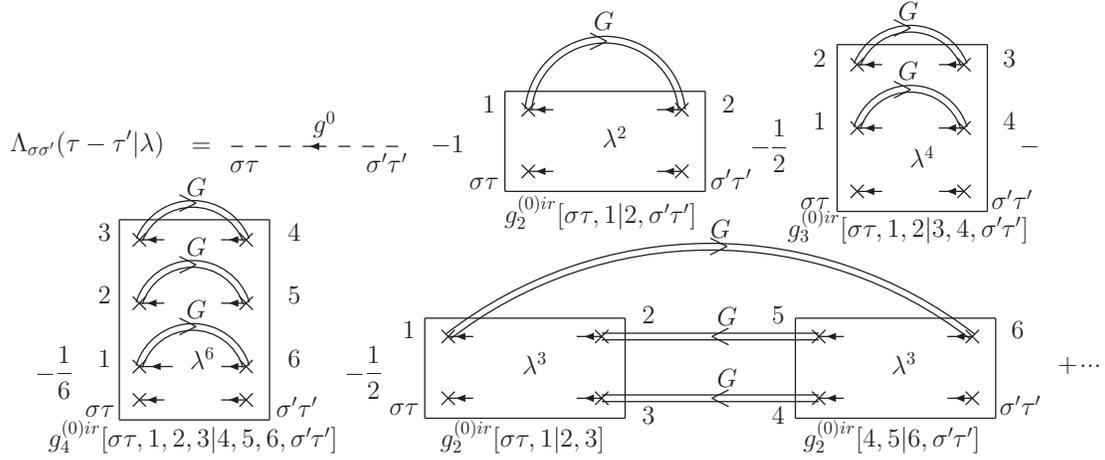}
\vspace{-0mm}
%
\caption{ Diagram representation of the correlation function
$\Lambda_{\sigma \sigma ^{\prime }}$. Double solid lines depict
the renormalized conduction electron propagators
$G_{\sigma \sigma ^{\prime }}(\tau -\tau ^{^{\prime }}|\lambda)$.
The arguments of irreducible functions are supposed arranged in the
clock wise direction. } \label{fig-3} \vspace{-5mm}
\end{figure*}
The series expansion of the full propagator $G_{\sigma \sigma^{\prime}}(\tau-\tau ^{\prime })$
can gives us more detailed representation of this quantity.

By using the Fourier representation of Matsubara functions on the
base of equations (9)and (10), we have:
%
\begin{equation}
%
G_{\sigma }(i\omega |\lambda)=\frac{G_{\sigma
}^{0}(i\omega )}{1-\Lambda _{\sigma }(i\omega |\lambda)G_{\sigma
}^{0}(i\omega )\lambda^{2}}.\label{12}
%
\end{equation}
%
\begin{equation}
%
g_{\sigma }(i\omega |\lambda)=\frac{\Lambda_{\sigma
}(i\omega |\lambda)}{1-\Lambda _{\sigma }(i\omega |\lambda)G_{\sigma
}^{0}(i\omega )\lambda^{2}}.\label{13}
%
\end{equation}
%
The equation (12) for conduction electron propagator is the Dyson
one with mass operator determined by the correlation function of
impurity electrons:
%
\begin{equation}
%
\Sigma_{\sigma }(i\omega |\lambda)=\lambda^{2}
\Lambda _{\sigma }(i\omega |\lambda).\label{14}
%
\end{equation}
%
When $\lambda$ is equal to one these quantities coincide.

The equation (13) for impurity electrons is of Dyson type and
coincides, for $\lambda=1$, with other equations obtained for
strongly correlated electrons $^{[17-21]}$. In equations (12),
(13) the parameter $\lambda$ can be taken equal to one and can be
omitted.


\section{Thermodynamic Potential}


The thermodynamic potential of our strongly correlated system is
equal to
%
\begin{eqnarray}
%
F&=&F_{0}-\frac{1}{\beta }\ln \left\langle U(\beta )\right\rangle
_{0},
\nonumber \\
&& \\
F_{0} &=&-\frac{1}{\beta }\ln Z_{0}-\frac{2}{\beta
}\sum\limits_{\mathbf{k} }\ln \left[ 1+\exp (-\beta \epsilon
(\mathbf{k)})\right],  \nonumber \label{15}
%
\end{eqnarray}
%

The diagrams which determine the mean value of the evolution
operator $\left\langle U(\beta )\right\rangle _{0}$ have not
external lines and are named vacuum diagrams. Between such
diagrams there are connected and disconnected ones. The
disconnected diagrams can be summed and the result of such
summation is equal to the exponent of connected diagrams. The
result of such summation permits us to formulate linked cluster
theorem. It has the form:
%
\begin{equation}
%
\left\langle U(\beta )\right\rangle _{0}=\exp \left\langle U(\beta
)\right\rangle _{0}^{c},\label{16}
%
\end{equation}
%
where $\left\langle U(\beta )\right\rangle _{0}^{c}$ is the infinite sum
of vacuum connected diagrams. This quantity is equal to zero
when hybridization is absent.

Therefore thermodynamic potential is equal to
%
\begin{equation}
%
F=F_{0}-\frac{1}{\beta }\left\langle U(\beta )\right\rangle
_{0}^{c},\label{17}
%
\end{equation}
%

In Fig.4 are depicted some of the simplest vacuum diagrams.
%
\begin{figure*}[t]
%
\centering
\includegraphics[width=0.85\textwidth,clip]{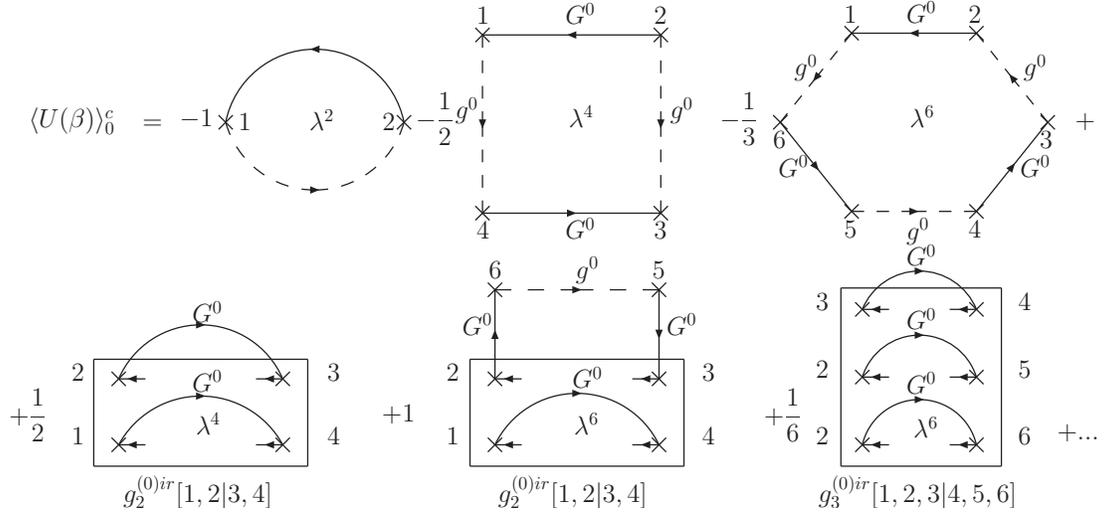}
\vspace{-0mm}
%
\caption{ Connected vacuum diagrams of the second, fourth and sixth
order of perturbation theory. } \label{fig-4} \vspace{-5mm}
\end{figure*}
%
The first three diagrams are of chain type and are originated from
the ordinary Wick contributions. The last three diagrams contain
the correlation functions and are determined by the new
contributions of GWT. The factor $\frac{1}{n}$, where $n$ is the
perturbation theory order, present in these diagrams makes it
difficult to carry out the summation over $n$. As is usual in such
cases $^{[23]}$ we employ a trick, that of integrating over the
interacting strength $\lambda$. The result of this procedure is
depicted in Fig.5.
%
\begin{figure*}[t]
%
\centering
\includegraphics[width=0.85\textwidth,clip]{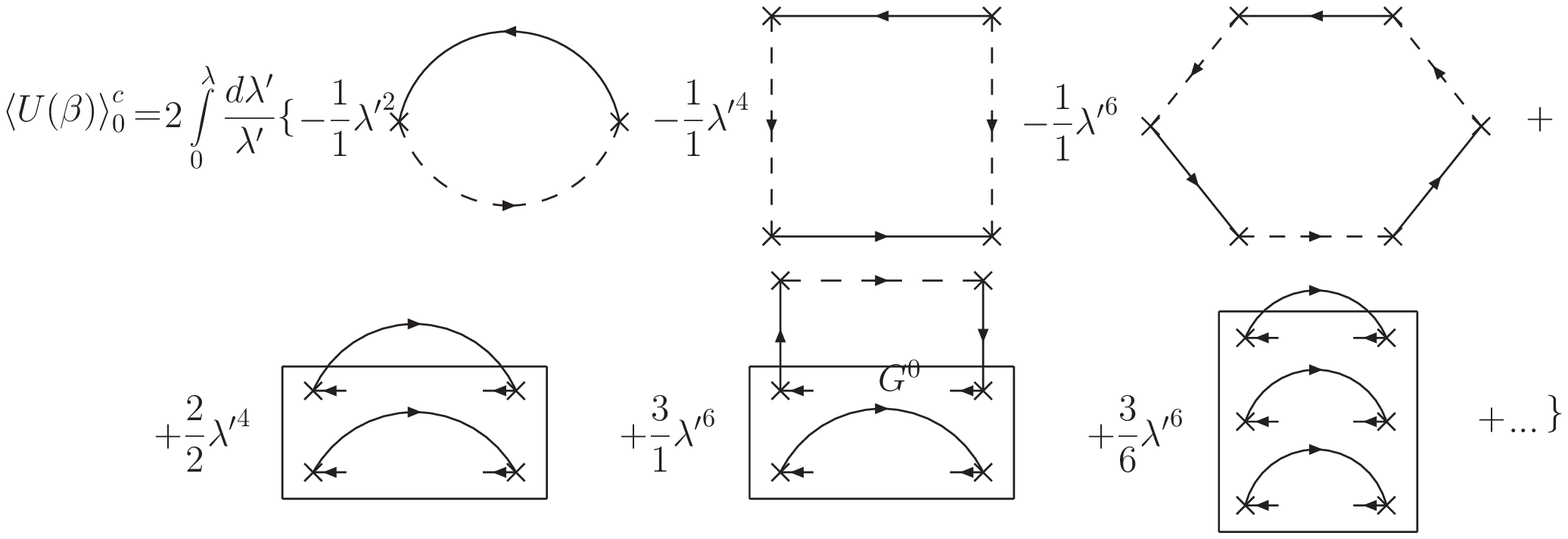}
\vspace{-0mm}
%
\caption{ The result of integration over the interacting strength
of the vacuum diagrams. } \label{fig-5} \vspace{-5mm}
\end{figure*}
%
Now we shall use the diagrams of the conduction electron
propagator $G_{\sigma}$ depicted on the Fig.1 and these of
$\Lambda_{\sigma}$ from Fig.3 to combine them in such a way to
obtain the vacuum diagrams of Fig.5. Either diagrams of Fig.5 of
the $n$ order of perturbation theory can be considered as the
product of the contribution of order $n_{1}$ from $G_{\sigma}$ and
of the contribution of order $n_{2}$ from $\Lambda_{\sigma}$ with
the condition that $n_{1}+n_{2}=n$. There are in general case
different possibilities to arrange such contribution and the
number of these possibilities is determined by the numerator of
the fraction before the diagrams of Fig.5. The denominator of this
fraction is determined from Fig.4. Therefore we obtain
\begin{widetext}
%
%
\begin{eqnarray}
%
\left\langle U(\beta )\right\rangle_{0}^{c}=(-2)\sum\limits_{\sigma _{1}\sigma_{2}}
\int\limits_{0}^{\beta }d\tau _{1}\int\limits_{0}^{\beta
}d\tau _{2}\int\limits_{0}^{\lambda}\frac{d\lambda ^{\prime }}{\lambda ^{\prime }}
G_{\sigma _{1}\sigma _{2}}(\tau _{1} -\tau _{2}|
\lambda ^{\prime })\Sigma_{\sigma _{2}\sigma _{1}}
(\tau _{2} -\tau _{1}|\lambda ^{\prime })=\\
=(-2)\sum\limits_{\sigma}\sum\limits_{\omega}\int\limits_{0}^{\lambda}
\frac{d\lambda ^{\prime }}{\lambda ^{\prime }}G_{\sigma }(i\omega |\lambda ^{\prime })
\Sigma_{\sigma }(i\omega |\lambda ^{\prime }).  \nonumber \label{18}
%
\end{eqnarray}
%
\end{widetext}

The thermodynamical potential becomes equal to
%
\begin{equation}
%
F=F_{0}+\frac{2}{\beta }\sum\limits_{\sigma}\sum\limits_{\omega}\int\limits_{0}^{\lambda}
\frac{d\lambda ^{\prime }}{\lambda ^{\prime }}G_{\sigma }(i\omega |\lambda ^{\prime })
\Sigma_{\sigma }(i\omega |\lambda ^{\prime }). \label{19}
%
\end{equation}
%
From this equation we obtain:
%
\begin{equation}
%
\lambda \frac{\partial F}{\partial \lambda}=\frac{2}{\beta }\sum\limits_{\sigma}\sum\limits_{\omega}
G_{\sigma }(i\omega |\lambda )
\Sigma_{\sigma }(i\omega |\lambda ). \label{20}
%
\end{equation}
%
The expression (19) for thermodynamical potential contains
additional integration over the interaction strength $\lambda$ and
is awkward because of it. As was proved for non correlated
many-electron system by Luttinger and Ward $^{[23]}$ this
expression can be transformed into a much more convenient formula.

We consider the following expression:
\begin{widetext}
%
%
\begin{eqnarray}
%
Y=-\frac{1}{\beta }\sum\limits_{\sigma}\sum\limits_{\omega}\exp(i\omega 0^{+})
\{\ln[G_{\sigma }^{0}(i\omega )\Sigma_{\sigma }(i\omega |\lambda ) -1]
+G_{\sigma }(i\omega |\lambda )
\Sigma_{\sigma }(i\omega |\lambda )\}+ Y ^{\prime }, \label{21}
%
\end{eqnarray}
%
\end{widetext}
which is the generalization of the Luttinger-Ward equation for
strongly correlated systems. Here $Y ^{\prime }$ is the sum of closed
linked skeleton diagrams with full $G_{\sigma}$ function as a contribution
of conduction electron lines.

On the Fig.6 are depicted some of simplest skeleton diagrams.
%
\begin{figure*}[t]
%
\centering
\includegraphics[width=0.85\textwidth,clip]{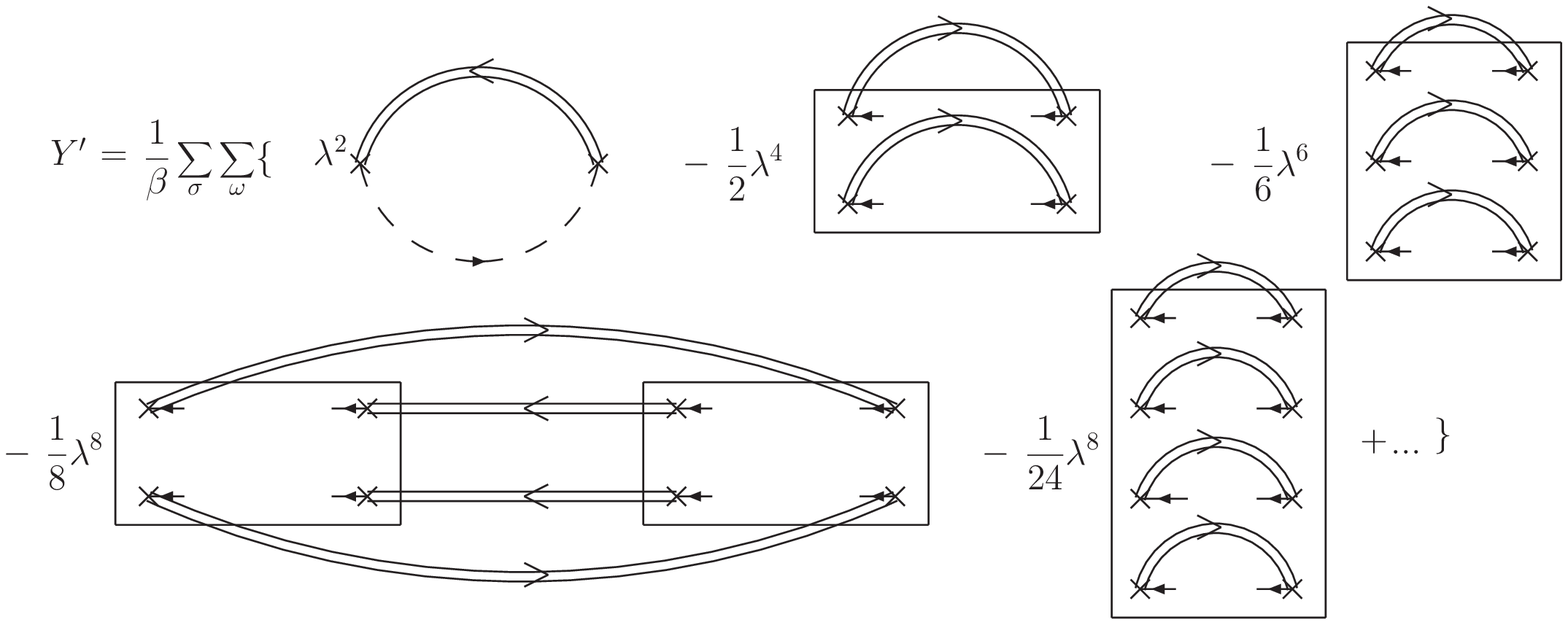}
\vspace{-0mm}
%
\caption{ Closed linked skeleton diagrams. The double solid lines
correspond to full propagators $G_{\sigma }(i\omega |\lambda )$ of
conduction electrons. The rectangles correspond to the correlation
functions of the correlated electrons. } \label{fig-6} \vspace{-5mm}
\end{figure*}
These diagrams depend on the interaction strength $\lambda$ not
only through the factors in front of each diagram but also through
the full Green's function $G_{\sigma }(i\omega |\lambda )$.

From equations (12), (14) and (21) we obtain
%
\begin{eqnarray}
%
\frac{\partial Y}{\partial \Sigma_{\sigma }(i\omega |\lambda )}&=
&-\frac{1}{\beta }\Sigma_{\sigma }(i\omega |\lambda )\
G_{\sigma }^{2}(i\omega |\lambda )
+\nonumber\\
+\frac{\partial Y^{\prime }}
{\partial \Sigma_{\sigma }(i\omega |\lambda )}, \label{22}
%
\end{eqnarray}
%
where, from Fig.3,6 and definition (14), it follows that
%
\begin{equation}
%
\frac{\partial Y^{\prime }}{\partial G_{\sigma }(i\omega |\lambda )}=
\frac{\lambda^{2}}{\beta }\Lambda_{\sigma }(i\omega |\lambda )=
\frac{\Sigma_{\sigma }(i\omega |\lambda )}{\beta }. \label{23}
%
\end{equation}
%
As a result we obtain the stationary property with respect to
changes of the mass operator:
%
\begin{equation}
%
\frac{\partial Y}{\partial \Sigma_{\sigma }(i\omega |\lambda )}=0. \label{24}
%
\end{equation}
%

Now we shall find the quantity $\frac{\partial
Y}{\partial\lambda}$ . By the stationary property of $Y$ we can
ignore the dependence of $\Sigma_{\sigma }$ and $G_{\sigma }$ on
$\lambda$ and take into account only the explicit dependence of
$\lambda$ in $Y^{\prime }$, depicted on the Fig.6. From this
figure it is easy to obtain:
%
\begin{equation}
%
\lambda \frac{\partial Y}{\partial \lambda}\mid_{\Sigma}=
\lambda \frac{\partial Y^{\prime}}{\partial \lambda}\mid_{\Sigma}=
\frac{2}{\beta }\sum\limits_{\sigma}\sum\limits_{\omega}
G_{\sigma }(i\omega |\lambda )
\Sigma_{\sigma }(i\omega |\lambda ). \label{25}
%
\end{equation}
%
From equations (20) and (25) we obtain
%
\begin{equation}
%
\lambda \frac{\partial F}{\partial \lambda}=
\lambda \frac{\partial Y}{\partial \lambda}. \label{26}
%
\end{equation}
%
The consequence of this equation is the solution
%
\begin{equation}
%
F(\lambda)=Y(\lambda)+\textsc{const}. \label{27}
%
\end{equation}
%
For $\lambda=0$ we have $Y(0)=0$ and $F(o)=F_{0}$. Therefore $\textsc{const}=F_{0}$.

The final result has the form
%
\begin{equation}
%
F=F_{0}+Y. \label{28}
%
\end{equation}
%


\section{Conclusions}


The thermodynamic potential of a strongly correlated system
described by the Anderson impurity model has been calculated. We
have formulated a new diagrammatic technique for fermions with
strong correlations and determined the correlation function of
localized electrons and mass operator of conduction electrons. For the
conduction electrons this operator coincides with the correlation
function of the impurity electrons. A Dyson's type of equation for
the one-particle propagators of both subsystems, of conduction and
impurity electrons, has been established. Within our diagrammatic
technique we first obtained an exact expression for the
thermodynamic potential as a product of the full propagator
$G_{\sigma}$ of the conduction electrons and its mass operator
$\Sigma_{\sigma}$, then a Luttinger-Ward-type $^{[23]}$ of identity based on the
stationary property of the potential was established. The
expression for the thermodynamic potential so obtained could be
very useful to calculate in a systematic way all thermodynamic
quantities (e.g. specific heat) of strongly correlated electron
systems.

%
\begin{acknowledgments}
We would like to thank Professor N.M. Plakida for a helpful discussion of
the paper. Two of us (V.M. and P.E.) thanks the Steering Committee of
the Heisenberg - Landau Program for support. One of us (V.M.) thanks
Theoretical Department of Duisburg - Essen University for hospitality and financial
support.

\end{acknowledgments}

%

%

\end{document}